# A Novel Method of Bolt Detection Based on Variational Modal Decomposition [1]

Juncai Xu[a,b], Qingwen Ren[a,2]

[a] Hohai University, Nanjing 210098, China
[b] Fundamental Science on Radioactive Geology and Exploration Technology Laboratory, East China Institute of Technology, Nanchang, Jiangxi, 330013, China

**Abstract**

The pull test is a destructive detection method, and it can't measure the actual length of the bolt. As such, ultrasonic echo is one of the most important non-destructive testing methods for bolt quality detection. In this paper, the variance modal decomposition (VMD) method is introduced into the bolt detection signal analysis. Based on the morphological filtering and the VMD method, the VMD combined morphological filtering principle (MF) is established into the bolt detection signal analysis method (MF-VMD). MF-VMD was used in order to analyze the simulation's vibration signal and the actual bolt detection signal. The results showed that the MF-VMD is able to effectively separate the intrinsic mode function, even when under the background of strong interference. Compared with the conventional VMD method, the proposed method is able to remove the noise interference. The intrinsic mode function of the field detection signal can be effectively identified by the reflection of the signal at the bottom of the bolt.

*Keywords:* bolt detection, variational modal decomposition, morphological filtering, intrinsic mode function.

## 1. Introduction

The bolt anchoring system is a hidden engineering process and is subject to geological conditions, construction technology and other environmental aspects[1]. As such, it is difficult to find hidden problems. The acoustic method is one of most important non-destructive testing methods used for bolt detection in civil engineering[2-4]. In order to obtain an effective signal, many data processing methods, such as the Short-time Fourier transform, Gabor transform, Wigner-Ville transform and wavelet transform, are proposed[5-7].

Wavelet transform has been adopted in many studies [8, 9]. However, the effect of wavelet transform is often limited by the selection of the wavelet bases and the decomposed layers. The empirical mode decomposition (EMD) can adaptively select the substrate according to the characteristics of the signal in multi-resolution, but it also avoids the selection of the wavelet basis[10, 11]. Nonetheless, EMD has the modal aliasing problem when processing the data.

The ensemble empirical mode decomposition (EEMD) was also proposed for solving the modal aliasing problem that is present on EMD[12]. However, because this method adds a different white noise, it may produce a false pattern after the decomposition, which may cause errors[13].

In recent years, the variational modal decomposition (VMD) method has been proposed[14]. This method transforms an input signal into several different constraint problems by Wiener filtering and Hilbert transform. It iterates the center frequency of each component and the bandwidth in order to achieve the adaptive decomposition of the signal. The results show that the method is superior to the traditional EMD method[15].

Based on the VMD theory, the VMD method is introduced into the bolt detection signal analysis. We

1) The project was supported by Fundamental Science on Radioactive Geology and Exploration Technology Laboratory（RGET1502）

2) E-mail：renqw@hhu.edu.cn

combined the principle of morphological filtering with VMD and proposed morphological filtering VMD (MF-VMD), and established the MF-VMD analysis method. The MF-VMD is used to simulate the vibration signal, as well as the actual Bolt detection signal processing in order to check its effect.

**2. Theory and methodology**

*2.1. Morphological Filter Principle*

Morphological filtering is achieved through the segmenting element moving in a signal to extract the signal's information, maintaining the details of the signal and removing the purpose of the noise interference. Morphological filtering is generally achieved by expansion, corrosion, opening and closing operations. Assuming that one vibration $s(n)$ has $N$ number of sampling points, then the segmenting element $g(m)$, $m=0,1,\ldots,M-1$, and the expansion and corrosion operations with $s(n)$ to $g(m)$ can be defined as follows:

$$(s \Theta g)(n) = \min_{m} \{s(n+m) - g(m)\} \quad (1)$$

$$(s \oplus g)(n) = \max_{m} \{s(n-m) + g(m)\} \quad (2)$$

Opening and closing operations with $s(n)$ to $g(m)$ can be defined as follows:

$$(s \circ g)(n) = (s \Theta g \oplus g)(n) \quad (3)$$

$$(s \bullet g)(n) = (s \Theta g \oplus g)(n) \quad (4)$$

In practical applications, the open-closed and the closed-and-open combined morphological filters are constructed through the use of the cascade form, which is used for noise reduction of the vibration signal. The expression is as follows:

$$MMC(s) = (s \circ g \bullet g + s \bullet g \circ g)/2 \quad (5)$$

The effect of the morphological filtering not only depends on the selected morphological operation, but it's also related to the structural elements that are used. The vibration signal is filtered by the linear structure's elements, as well as the correlation between the acoustic signal before and after the filtering is chosen as the criterion of the selection width value in the research.

*2.2. VMD Principle*

VMD is a variational problem. In order to minimize the sum of the estimated bandwidths of each mode, we assumed that each mode has a finite bandwidth with different central frequencies. As a result, the alternating direction multiplier method was adopted in order to constantly update the mode and its central frequency, with the mode being gradually demodulated to its corresponding baseband. Then, the final mode and the corresponding center frequency were extracted.

Assuming that a signal $S_0$ is decomposed into $N$ intrinsic mode function (IMF), then the corresponding variational problem's solution can be expressed as follows:

1) The Hilbert transform of each IMF component is used to obtain the analytic signal

$$\left(\delta(t)+\frac{j}{\pi t}\right)*u_k(t) \tag{6}$$

2) The center frequency is estimated with the obtained analytic signal and the spectrum of each analytical signal is transformed into the baseband with a frequency shift

$$\left[\left(\delta(t)+\frac{j}{\pi t}\right)*u_k(t)\right]e^{-j\omega_k t} \tag{7}$$

3) The $L^2$ norm of the demodulated signal is calculated, and the bandwidth of each mode is estimated. The variational problem is expressed as follows:

$$\begin{cases} \min_{\{u_k\},\{w_k\}}\left\{\sum_k\left\|\partial_t\left[\left(\delta(t)+\frac{j}{\pi t}\right)*u_k(t)\right]e^{-j\omega_k t}\right\|^2\right\} \\ s.t.\sum_k u_k = f(t) \end{cases} \tag{8}$$

4) For the mentioned variational problem, the quadratic penalty function and the Lagrange multiplier can be used in order to transform the problem into an unconstrained problem form,

$$L(\{u_k\},\{\omega_k\},\lambda) = \alpha\sum_k\left\|\partial_t\left[\left(\delta(t)+\frac{j}{\pi t}\right)*u_k(t)\right]e^{-j\omega_k t}\right\|_2^2 \\ + \left\|f(t)-\sum_k u_k(t)\right\|_2^2 + \left\langle\lambda(t), f(t)-\sum_k u_k(t)\right\rangle \tag{9}$$

where $\alpha$ is the penalty factor, and $\lambda(t)$ is the Lagrange multiplier.

Finally, the multiplier alternate direction algorithm is used in order to solve the unconstrained variational problem of equation (9), and then the IMF can be obtained.

### 3. Simulation Signal Analysis

The anchor's anchoring detection signal is regarded as a vibration signal. The singular point of the vibration signal is usually used to identify the anchoring quality mark of the bolt. One simulation signal was adopted in order to check the ability of the identifying abnormality of the VMD and the study recognition of the VMD and MF-VMD for the vibration signal in an environment with strong noise.

*3.1. The VMD decomposition of Simulation Signal*

One simulation signal $s(t)$ has a singular point and two frequencies, 10kHz and 20kHz. There are singular points at 0.8ms and at 1.2ms (Fig. 1).

$$s(t) = \begin{cases} \sin(20000\pi t), & 0\,\text{ms} < t \leq 0.8\,ms \\ \sin(40000\pi t), & 0.8\,\text{ms} < t \leq 1.2\,ms \\ \sin(20000\pi t), & 1.2\,\text{ms} < t \leq 2\,ms \end{cases} \tag{10}$$

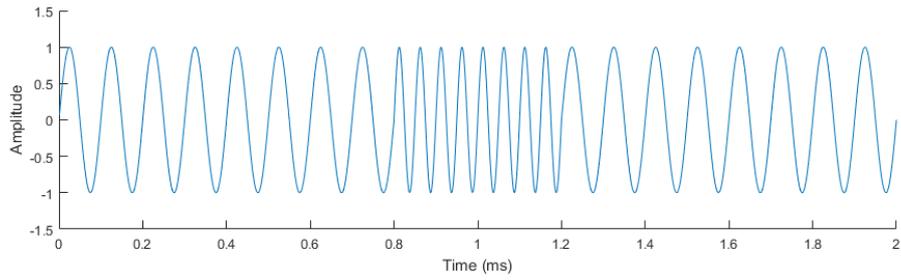

Fig.1. Simulation signal *s*(*t*)

The simulation test signal *s* (*t*) in Fig. 1 is decomposed by the VMD method. The result of the decomposition is shown in Fig.2.

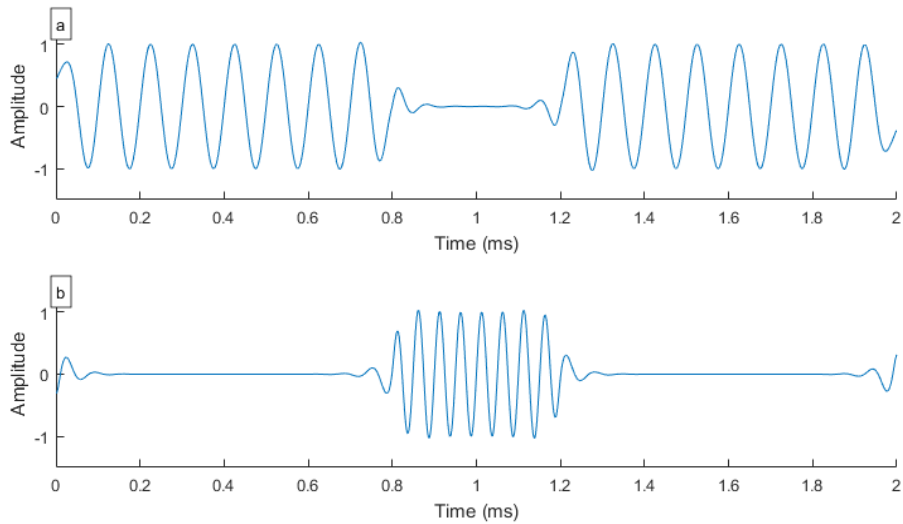

Fig.2. The decomposition result of VMD (a)IMF1 (b)IMF2

Based on Fig.2, IMF1 and IMF2 can both be completely obtained from signal *s*(*t*) and it can display two kinds of vibration. The instantaneous frequency spectrum is shown in Fig.3.

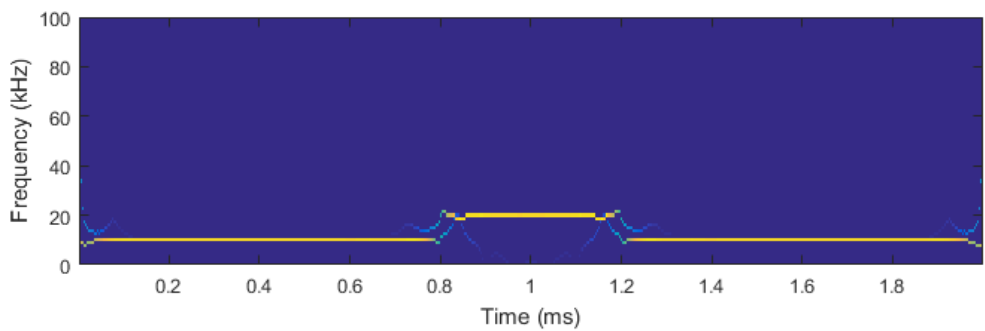

Fig.3. Instantaneous frequency with VMD

As can be seen in Fig.3, the two kinds of frequencies, 10kHz and 20kHz, appear in the signal. The singularity points are at 0.8ms and at 1.2ms. Thus, VMD is able to decompose the different modes from the signal when there are no noise interferences

## 3.2. Simulation Signal under Noise Interferences

In the studied vibration signal decomposition with VMD, $s(t)$ in Fig.1 was added with noise when SNR=5dB. The simulation signal $s_n(t)$ that contains the noise interference is shown in Fig.4.

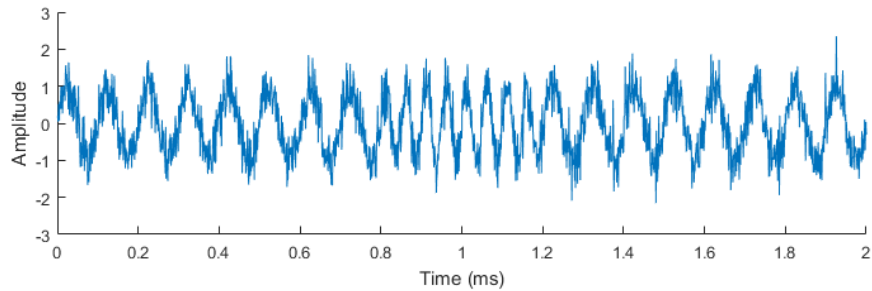

Fig.4. $s_n(t)$ containing the noise interference

We use the VMD to decompose $s_n(t)$. The different IMF can be obtained with VMD. The results are shown in Fig.5.

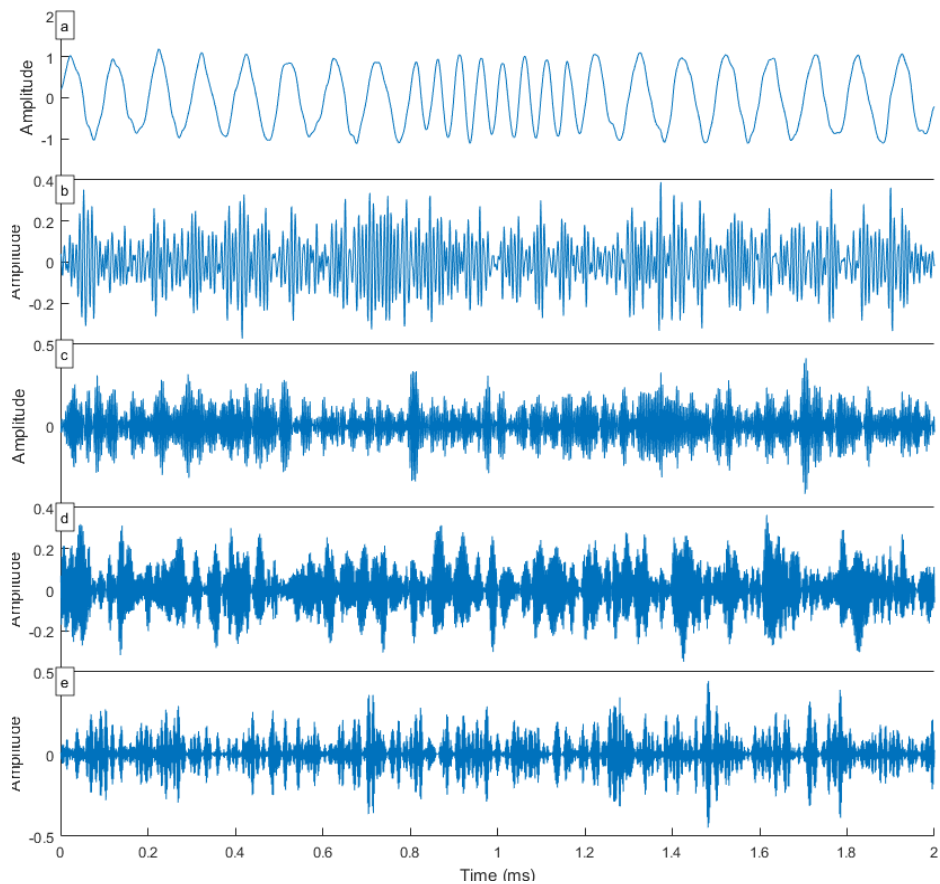

Fig.5. Decomposition results of VMD (a) IMF1 (b) IMF2 (c) IMF3 (d) IMF4 (e) IMF5

Based on Fig.5, IMF1 has a small difference with the original signal $s(t)$. However, IMF2~5 are noise. VMD is unable to decompose the different modes from $s_n(t)$, as there are two types of mode signal mixing in IMF1. The Hilbert instantaneous frequency comes from VMD in Fig.6.

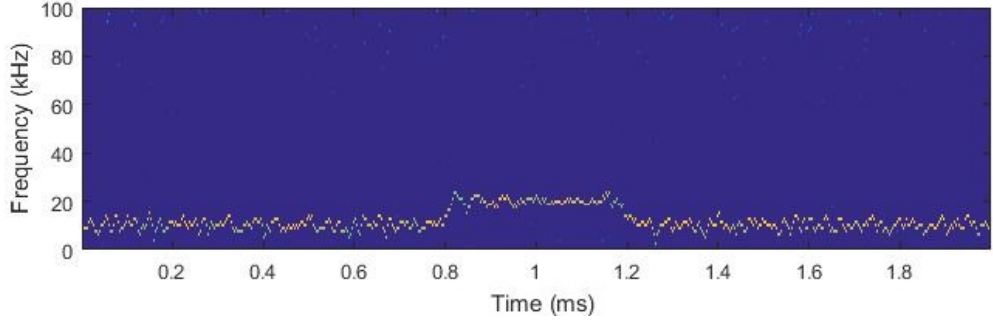

Fig.6. Hilbert instantaneous frequency

As can be seen in Fig.6, the instantaneous frequency is continuous from the VMD method. There are no significant singularities in the instantaneous frequency, so the vibration signal has obvious interference.

MF-VMD was also adopted in order to decompose $s_n(t)$. The IMFs of $s_n(t)$ can be obtained, as is shown in Fig.7.

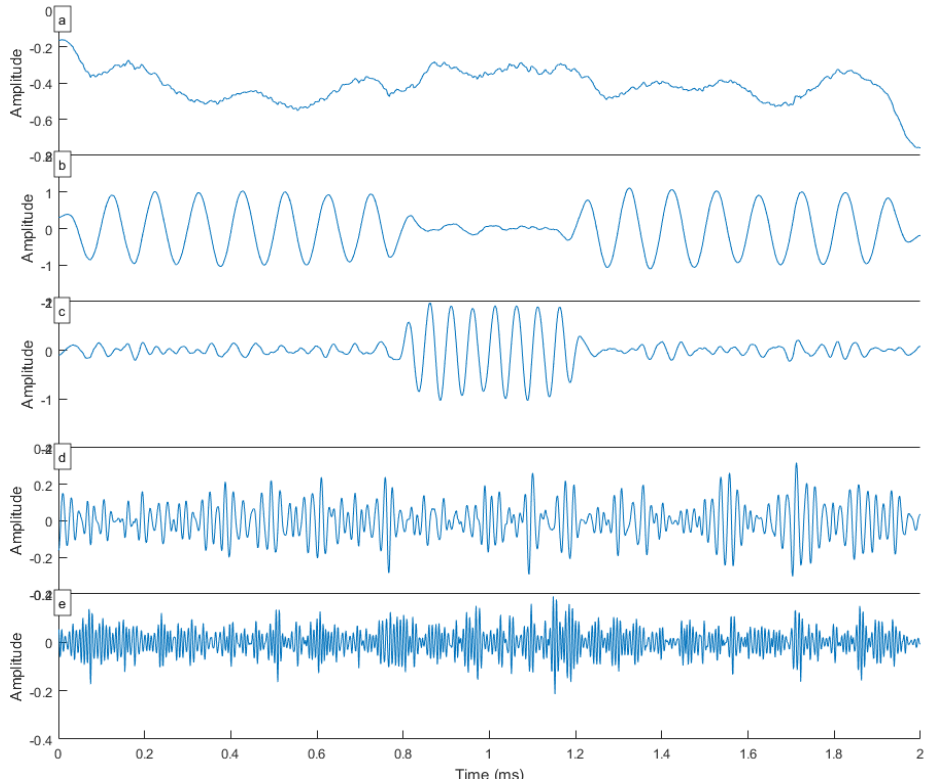

Fig.7. MF-VMD decomposition results (a) IMF1　(b)IMF2　(c)IMF3　(d)IMF4　(e)IMF5

As is shown in Fig.7, we are able to find that IMF1, IMF4 and IMF5 are the residual noise signals, and the IMF2, IMF3 are the two different modes of s (t), and correspond to IMF1 and IMF2 in Fig.2. The IMF was taken from the Hilbert transformation; the Hilbert instantaneous spectrum is shown in Fig.8.

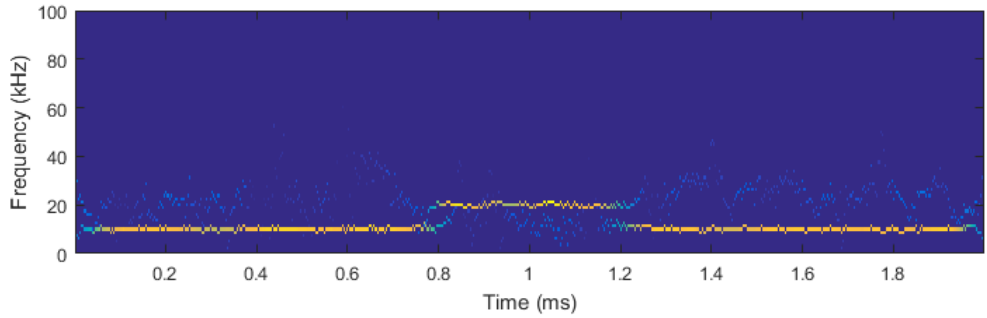

Fig.8. Hilbert instantaneous frequency

As can be seen from the spectrum in Fig.8, the instantaneous frequency with the MF-VMD decomposition has some obvious singularities at both frequencies of 10 kHz and 20 kHz. The random interference signal is well suppressed when compared with the spectrum in Fig. 6. The frequency strength's display is significantly higher than the frequency strength that is seen in Figure 6. In a strong noise background, MF-VMD has good recognition of the singular point of the vibration signal.

## 4. Bolt Anchoring Detection Signal Analysis

Taking the high-slope bolt anchoring testing site in Yunnan expressway in China as an example, the instrument used is the AGI-MG bolt quality detector (Fig. 9). The sampling parameter is 1.05kHz, the sampling number is 980, and the sampling interval is 4.0us. The collected vibration signal is shown in Fig.10.

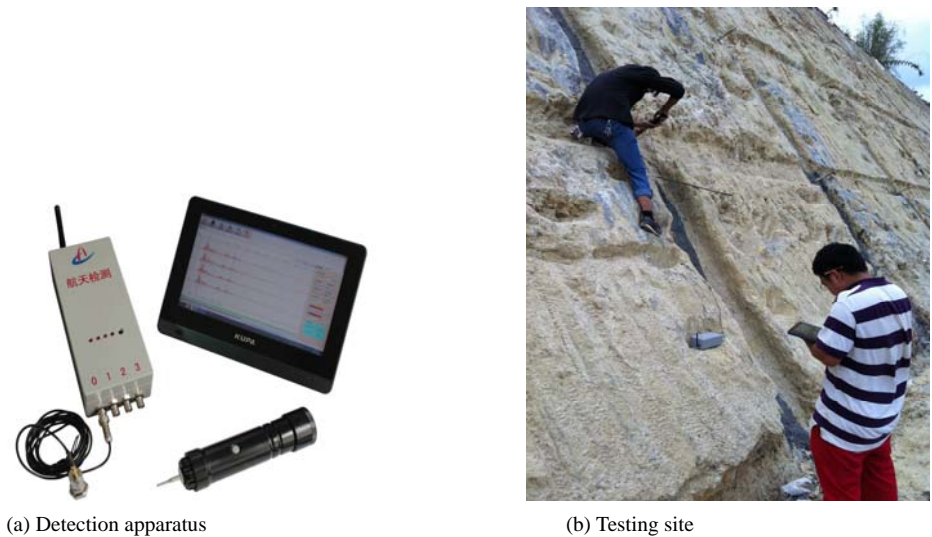

(a) Detection apparatus          (b) Testing site

Fig.9. Bolt anchoring testing site

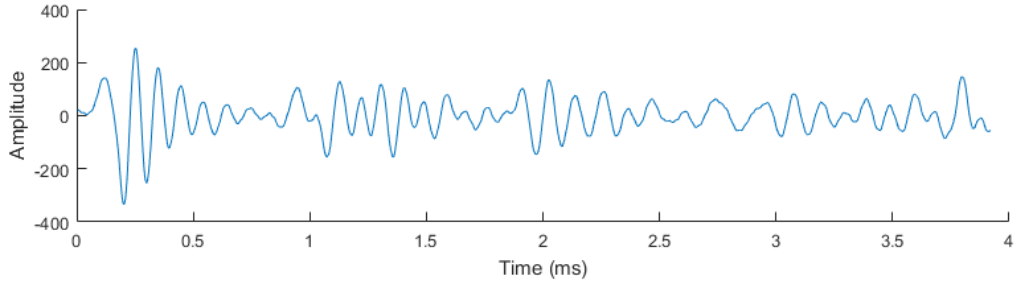

Fig.10. bolt anchoring detection signal

It is difficult to directly determine the bottom of the bolt reflection signal in Fig.10. As such, we adopted MF-VMD in order to decompose the signal. The decomposition result is shown in Fig.11.

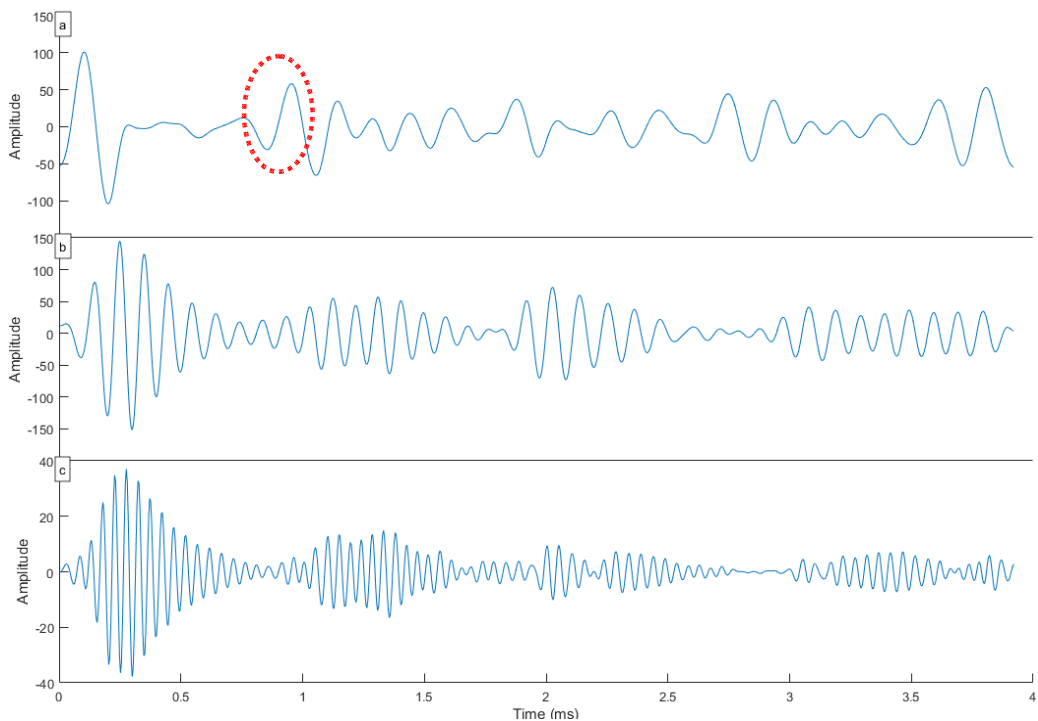

Fig.11. Decomposition results with MF-VMD (a) IMF1　(b) IMF2　(c) IMF3

According to Fig.11, three IMFs were obtained by MF-VMD. There is one obvious reflection signal from the bolt bottom at 1.0ms in IMF1. As the bolt length is 3, and the velocity of the wave in the bolt is 6000m/s, then the reflection signal should be at 1.0ms. Thus, the proposed method is able to decompose the reflection signal of the bolt bottom and determine the reflection time.

## 5. Conclusions

Based on the VMD principle, the VMD theory was introduced into the detection signal analysis of bolt anchoring. With the combination of MF and VMD, we proposed the MF-VMD method to analyze the bolt detection signal. Based on the analysis of the simulation signal, as well as the field application, we are able to draw some conclusions, which are the following:

1）VMD is able to decompose the different modes from the vibration signal. However, VMD can't properly decompose the IMFs from vibration when there is strong noise interference.

2）MF-VMD is able to properly decompose the IMFs from vibration signal, even when under strong noise interference and reduce the effect of noise.

3）MF-VMD is able to properly decompose the bolt detection signal into the IMFs, and the reflection signal from the bottom of the bolt can also be identified in IMF.

**Acknowledgments**

This research was funded by the Open Research Fund of the Fundamental Science on Radioactive Geology and Exploration Technology Laboratory (Grant No.RGET1502) and A Project Funded by the Priority Academic Program Development of Jiangsu Higher Education Institutions (Grant No.3014-SYS1401).